\documentclass[nofootinbib,11pt,
      twoside,preprintnumbers,notitlepage]{revtex4}
\usepackage{epsfig,amsmath,amssymb}
\usepackage{color}
\usepackage{float}
\usepackage[colorlinks]{hyperref}
\usepackage{subfigure}
\usepackage[utf8]{inputenc}
\usepackage{slashed}
\usepackage{times}
\usepackage{epstopdf}
\usepackage{booktabs}
\usepackage{multirow}
\usepackage[normalem]{ulem}
\usepackage[T1]{fontenc} 

\hypersetup{
    linktocpage,
     colorlinks,
     citecolor=darkgreen,
     linkcolor= darkgreen,
     urlcolor=darkgreen
}
\definecolor{darkred}{rgb}{0.5,0.0,0.0}
\definecolor{darkblue}{rgb}{0.0,0.0,0.9}
\definecolor{darkerblue}{rgb}{0.0,0.0,0.5}
\definecolor{purple}{rgb}{0.5,0.0,0.5}
\definecolor{darkgreen}{rgb}{0.0,0.5,0.0}
\definecolor{black}{rgb}{0.0,0.0,0.0}
\definecolor{brown}{rgb}{0.6,0.4,0.2}
\definecolor{newpurple}{rgb}{0.65, 0.38, 0.61}
\definecolor{newyellow}{rgb}{0.9718, 0.6093, 0.0759}
\definecolor{amber}{rgb}{1.0, 0.75, 0.0}
\definecolor{newblue}{rgb}{0.4, 0.52, 0.85}
\definecolor{newred}{rgb}{0.8524, 0.2595, 0.3294}
\definecolor{newgreen}{rgb}{0.2, 0.8, 0.2}
\definecolor{SMgreen}{rgb}{0.56, 0.69, 0.19}
\definecolor{neworange}{rgb}{0.94, 0.462, 0.162}

\definecolor{BrickRed}{rgb}{0.9,0.1,0}

\newcommand{\bea}{\begin{eqnarray}}
\newcommand{\eea}{\end{eqnarray}}
\newcommand{\beq}{\begin{equation}}
\newcommand{\eeq}{\end{equation}}
\newcommand{\ec}{\end{center}}
\newcommand{\bc}{\begin{center}}

\newcommand{\eq}[1]{Eq.~(\ref{#1})}

\begin{document}

\preprint{TTP-21-028, P3H-21-060, RBI-ThPhys-2021-34}
\title{Higgs decay into a lepton pair and a photon:\\ a roadmap  to $\mathbf{H\to
  Z\gamma}$ discovery and probes of new physics}
\author{Aliaksei Kachanovich$^1$, Ulrich Nierste$^1$, and Ivan  Ni\v sand\v zi\'c$^2$} 

\date{\today}

\email[Electronic addresses:]{aliaksei.kachanovich@kit.edu,
 ulrich.nierste@kit.edu, ivan.nisandzic@irb.hr}

\affiliation{$^1$Institut f\"ur Theoretische Teilchenphysik (TTP),
 Karlsruher Institut f\"ur Technologie (KIT), 76131 Karlsruhe,
 Germany.\\
$^2$Ru\dj er Bo\v skovi\'c Institute, Bijeni\v cka cesta 54, 10000, Zagreb, Croatia.}

\begin{abstract}
  The decay $H\to \ell^+\ell^- \gamma$, $\ell=e,\mu$, receives
  contributions from $H\to Z[\to \ell^+\ell^-] \gamma$ and a
  non-resonant contribution, both of which are loop-induced. We describe
  how one can separate these sub-processes in a gauge-independent way,
  define the decay rate $\Gamma(H\to Z\gamma)$, and extract the latter
  from differential $H\to \ell^+\ell^- \gamma$ branching ratios.  For
  $\ell=\mu$ also the tree decay rate, which is driven by the muon
  Yukawa coupling, is important. We propose kinematic cuts optimized to
  separate the three contributions, paving the way to the milestones (i)
  discovery of $H\to Z  \gamma$, (ii) discovery of
  $H\to \mu^+\mu^- \left. \gamma\right|_{\rm tree}$, and (iii)
  quantification of new physics in both the effective
  $H$-$Z$-$\gamma$ and 
  non-resonant $H$-$\ell^+$-$\ell^-$-$\gamma$ couplings.
\end{abstract}

\maketitle

\section{Introduction}
Currently ATLAS and CMS put substantial effort into the discovery of the
decay $H\to Z\gamma$.  However, this process is only well-defined when
the $Z$ boson is taken on-shell. If one includes the effect of a
non-vanishing $Z$ decay width $\Gamma_Z$ by smearing the off-shell
$H\to Z\gamma$ decay amplitude with a Breit-Wigner distribution one
finds an unphysical, gauge-dependent result \cite{Passarino:2013nka}.
If the $Z$ boson is detected through its leptonic decay, $H\to Z\gamma$
is a sub-process of $H \to \ell^+\ell^- \gamma$. The one-loop diagrams
contributing to the process $H \to \ell^+\ell^- \gamma$ can be
divided into three classes, namely diagrams with off-shell Z boson
(describing $H\to Z^\ast [\to \ell^+ \ell^-]\gamma $), those with
off-shell photon (involving
$H\to \gamma^\ast [\to \ell^+ \ell^-]\gamma$), and box-diagrams. The
calculations of the $H \to\ell^+\ell^- \gamma$ decay amplitude in an
arbitrary linear $R_\xi$-gauge in Ref.~\cite{Kachanovich:2020xyg} has
revealed how the sum of all diagrams in each class depend on the gauge
parameter $\xi$ of the $W$ boson.  This dependence cancels in the
final physical result after the summation of all contributions. Complete
one-loop calculations of differential decay rates (and asymmetries) of
$H \to\ell^+\ell^- \gamma$ in the Standard Model (SM) have been performed by
several groups
\cite{Passarino:2013nka,Kachanovich:2020xyg,Abbasabadi:1996ze,Dicus:2013ycd,Han:2017yhy}
and Ref.~\cite{Kachanovich:2020xyg} contains a detailed comparison of
the  numerical results presented in these references.

Nevertheless, it is possible to define a gauge-independent resonant
contribution which peaks near $s=M_Z^2$, where $\sqrt{s}$ is the
invariant lepton mass. The remaining contribution to
$H \to \ell^+\ell^- \gamma$, consisting of
$H\to \gamma^\ast [\to \ell^+ \ell^-]\gamma$, box diagrams, and the
gauge-dependent off-peak pieces of
$H\to Z^\ast [\to \ell^+ \ell^-]\gamma $ are all non-resonant and can be
experimentally distinguished from the resonant term of interest. Next
one can employ the \emph{narrow-width approximation (NWA)}\ to relate
the latter to the product of the decay rate $\Gamma(H\to Z\gamma)$ and
the branching ratio $BR(Z\to \ell^+\ell^-)$. Thus one arrives at a
physical, experimentally accessible definition of
$\Gamma(H\to Z\gamma)$. Then ruling out $\Gamma(H\to Z\gamma)=0$ at five
standard deviations will constitute the desired discovery of this decay
mode\footnote{$t\text{-}\bar{t}$-associated Higgs production has
  been recently proposed \cite{Goertz:2019uek} as a particularly
  promising channel for the discovery of $H\to Z\gamma$ at the
  high-luminosity LHC.}. At several steps of this derivation (for
instance by modifying the NWA) one could change the definition of
$\Gamma(H\to Z\gamma)$ by terms of order $\Gamma_Z^2/M_Z^2$ and arrive
at equally valid, yet different results. This feature is intrinsic to
any decay into an unstable particle detected only through its decay
products. In view of the smallness of $\Gamma_Z^2/M_Z^2\sim 10^{-3}$,
however, this ambiguity is phenomenologically irrelevant.

The differential decay rate  $d\Gamma (H \to \mu^+\mu^- \gamma)/dm_{\mu\mu}$  peaks
at the photon and $Z$ poles at $m_{\mu\mu}=0$ and $m_{\mu\mu}\simeq M_Z$, respectively,
and rises towards the end of the spectrum at $m_{\mu\mu}=M_H$ (see
Fig.~\ref{Fig:Distributions} (b)).  The latter effect is due to the
tree-level contribution involving the small muon Yukawa coupling.  ATLAS
has already found evidence for $H \to \ell^+\ell^- \gamma$ in the low
invariant mass region dominated by the photon pole \cite{ATLAS:2021wwb}.
To discover $H\to Z\gamma$ one must study the complementary region and
in the $H \to \mu^+\mu^- \gamma$ data carefully separate the $Z$ peak
from $H\to \mu^+\mu^- \left. \gamma\right|_{\rm tree}$. A discovery of
the latter contribution will constitute a manifestation of the Higgs
Yukawa coupling to muons, independent of and complementary to the
observation of $H\to \mu^+\mu^-$.  The loop contribution to the decay
rate of $H\to e^+ e^- \gamma$ is several orders of magnitude larger than
the corresponding tree contribution, as the latter is suppressed by the
square of the tiny electron Yukawa coupling.  We do not consider the
process $H\to \tau^+\tau^-\gamma$ which is dominated by the tree-level
contribution. The light lepton masses are neglected in the loop
contributions which are found infrared-finite in this limit.

With increasing statistics one will be able to quantify deviations from
the SM predictions not only for the effective $H$-$Z$-$\gamma$ vertex,
but also for the effective non-resonant $H$-$\ell^+$-$\ell^-$-$\gamma$
couplings.  To this end the data sample with $\ell=e$ and $\ell=\mu$
should not be combined, as new-physics (NP) contributions are likely to
be different. Through the Higgs vev $H$-$\mu^+$-$\mu^-$-$\gamma$
couplings can contribute to the anomalous magnetic moment of the muon,
whose measurement significantly deviates from the SM prediction
\cite{Muong-2:2021ojo}. The non-resonant region between photon and $Z$
pole is best suited to probe those NP operators which are unrelated to
the effective $H$-$Z$-$\gamma$ vertex, because the SM contribution is
small.
  
This paper is organized as follows: In Sec.~\ref{Sec:res} we separate the
gauge-independent resonant contribution to $H \to \ell^+\ell^- \gamma$
related to the $H\to Z \gamma$ sub-process. Sec.~\ref{Sec:Kinematic cuts}
proposes various kinematic cuts to enhance the sensitivities to  $H\to Z
\gamma$, $H\to  \mu^+\mu^- \left. \gamma\right|_{\rm tree}$, or
non-resonant NP. 
In Sec.~\ref{Sec:nwa} we define $B(H\to Z\gamma)$ and relate this
quantity to the resonant piece of $H \to \ell^+\ell^- \gamma$ and
Sec.~\ref{Sec:con} presents the conclusions. Two appendices contain
numerical input values and the loop function for $H\to Z \gamma$.

\section{Separating the resonant contribution\label{Sec:res}}

We parametrize the loop-induced amplitude for the process $h\to \ell\ell\gamma$ as:
\begin{eqnarray}
  \mathcal{A}_{\text{loop}}&=&\big[(k_\mu\,p_{1\nu}-g_{\mu\nu}\,
                                k\cdot p_1)  \bar{u}(p_2)\big(a_1
                                \gamma^\mu P_R 
   + b_1 \gamma^\mu P_L\big) v(p_1)\nonumber\\
                            &+&(k_\mu\,p_{2\nu}-g_{\mu\nu}\, k\cdot p_2)
                                \bar{u}(p_2)
 \big(a_2\gamma^\mu P_R + b_2\gamma^\mu P_L\big)v(p_1)\big]
 \varepsilon^{\nu\,\ast}(k)\,,
\label{loop_amp}
\end{eqnarray}
where, using the notation of Ref.~\cite{Kachanovich:2020xyg}, we denote
the four-momenta of photon, lepton and antilepton by $k$, $p_1$, $p_2$,
respectively, while the chiral projectors are
$P_{L,R}=(1\mp\gamma_5)/2$.

The loop-functions $a_{1,2}$ and $b_{1,2}$ depend on the Mandelstam
variables
\begin{equation}
  s=(p_1+p_2)^2, \qquad t=(p_1+k)^2,\qquad  \mbox{and}\quad
  u=(p_2+k)^2 =m_H^2+2m_\ell^2-s-t, \label{eq:man}
\end{equation}
where $m_\ell$ and $m_H$ denote the masses of lepton and Higgs
boson. The coefficients $a_2$ and $b_2$ are obtained by exchanging the
variables $t$ and $u$ within $a_1$ and $b_1$, respectively. Explicit
one-loop expressions for the coefficients $a_1$ and $b_1$ can be found
in Ref.~\cite{Kachanovich:2020xyg} and corresponding ancillary files.

Each of the coefficients $a_{1,2}$ and $b_{1,2}$ can be written in the following form, e.g. for $a_1$:
\begin{eqnarray}
a_{1}(s,t)&=&\widetilde{a}_1(s,t)+\frac{\alpha_{1}(s)}{s-m_Z^2+i m_Z \Gamma_Z}\,,\label{Eq:separating}
\end{eqnarray} 
with the obvious index replacement and the change of notation
$\alpha_{1,2} \to \beta_{1,2}$ for the coefficients $b_{1,2}$. Note the
relations
\begin{equation}
\alpha_1(s) = \alpha_2(s) \equiv \alpha(s)\qquad  \mbox{and}\quad
\beta_1(s) = \beta_2(s) \equiv \beta(s) . \label{eq:rel}
\end{equation}
As mentioned in the
Introduction, the off-shell amplitude for $H\to \gamma Z^\ast$, which
determines $\alpha(s)$ and $\beta(s)$, depends on the unphysical
gauge parameter $\xi$.  However, the process $H\to \gamma Z$ involving
the on-shell Z boson does not depend on the gauge. Thus, we can isolate
the $\xi$-independent part of the amplitude for
$H\to \gamma Z^\ast [\to \ell^+ \ell^-]$ sub-process by setting
$s=m_Z^2$ in $\alpha(s)$, $\beta (s)$, {\it i.e.}\ the residue
of the Z-boson propagator is gauge-independent. In the following we
denote this term the "resonant" contribution.

Separating the resonant and non-resonant terms in this way yields
\beq
a_1(s,t) = a_1^{nr}(s,t) + a_1^{res}(s)\,,\label{Eq:separating2}
\eeq 
\begin{eqnarray}
a_1^{nr}(s) \equiv \widetilde{a}_1(s,t) + \frac{\alpha(s)-\alpha(m_Z^2)}{s-m_Z^2+i m_Z \Gamma_Z}\,,\qquad\qquad
a_1^{res}(s) \equiv \frac{\alpha (m_Z^2)}{s-m_Z^2+i m_Z \Gamma_Z}. \label{eq:a1}
\end{eqnarray}   
We write
\begin{eqnarray}
  \frac{d^2\Gamma}{ds\,dt}&=&  \frac{d^2\Gamma_{\text{loop}}}{ds\,dt} +
                              \frac{d^2\Gamma_{\text{tree}}}{ds\,dt},  \nonumber
\end{eqnarray}
where the tree contribution in the second term is to be dropped for $\ell=e$. 
The loop contribution to the differential decay rate over the variables $s$ and $t$ is given by the formula:
\begin{equation}
\frac{d^2\Gamma_{\text{loop}}}{ds\,dt}=\frac{s}{512 \pi^3 m_H^3}\big[t^2(\vert a_1\vert^2 + \vert b_1\vert^2)+u^2 (\vert a_2\vert^2+\vert b_2\vert^2)\big]\,,
\label{loop-decayrate}
\end{equation}
where we have neglected the light lepton masses in the phase space and
$u$ is to be substituted for the expression in \eq{eq:man}. The
non-zero value of the lepton mass impacts the value of the loop induced
contribution to the decay rate only in the dilepton invariant-mass
region close to the production threshold, $m_{\ell\ell} \sim 2 m_\ell$,
via the kinematic effect. We avoid this region by using the cut
$m_{\ell\ell\,, min}\equiv \sqrt{\tilde
    s_{\rm  min}} = 0.1\,m_\text{H}$ in what follows.

The square of the magnitude of $a_1$ in \eq{eq:a1} contains three distinguishable pieces:
\begin{eqnarray}
  \vert a_1\vert ^2 =\; \vert a_1^{nr}\vert ^2 \;+\;  \vert a_1^{res}\vert ^2
                         \;+\;2\, \mbox{Re}\,(a_1^{nr} a_1^{res^\ast})\, ,\label{Eq:squares}
\end{eqnarray}
and \emph{mutatis mutandis} for $a_2$ and $b_{1,2}$.
Corresponding contributions to the one-loop decay rate are
\begin{eqnarray}
  \frac{d^2\Gamma_{\text{loop}}}{ds\,dt} = \frac{d^2\Gamma_{nr}}{ds\,dt}
       +  \frac{d^2\Gamma_{res}}{ds\,dt} +  \frac{d^2\Gamma_{int}}{ds\,dt}\,,
\end{eqnarray}
where the small interference term, denoted by $\Gamma_{int}$, corresponds to the
third term in Eq.~\eqref{Eq:squares} and can be safely neglected for the
purposes of expected near-future measurements. 

The differential decay rate for the tree contribution for $H \to \mu^+\mu^-\gamma$ reads:
\begin{eqnarray}
  \displaystyle \frac{d^2\Gamma_{\text{tree}}}{ds\,dt}
  &=& \mathcal{N}\Big[\frac{9 m_\mu^4+\;m_\mu^2(-2s +t -3 u)+\; t\,u}{(t-m_\mu^2)^2}+
      \frac{9 m_\mu^4+\;m_\mu^2(-2s +u -3 t)+\; t\,u}{(u-m_\mu^2)^2}\nonumber \\
  &&\qquad
     +\; \frac{34 m_\mu^4-\; 2 m_\mu^2\left(8s+5(t+u)\right)+\; 2(s+t)(s+u)}{(t-m_\mu^2)(u-m_\mu^2)}\Big]\, ,\label{Eq:tree-decayrate}
\end{eqnarray}
where
\begin{equation}
\label{norm}
\mathcal{N} =  \frac{e^{4} m_{\mu}^{2}}{256\,\pi^{3} \sin^{2}\theta_{W} m_{W}^{2} m_{H}^{3}}\,.
\end{equation}
For this distribution, we keep the nonvanishing muon mass in the
formulas for physical kinematic limits given in
Eq.~\eqref{Eq:t-range}. Note that the muon mass cannot be neglected
in the phase space integral of the tree contribution, see
Eq.\eqref{Eq:t-range} below.

\begin{figure}[t]
	\begin{center}
		\subfigure[t][]{\includegraphics[width=0.485\textwidth]{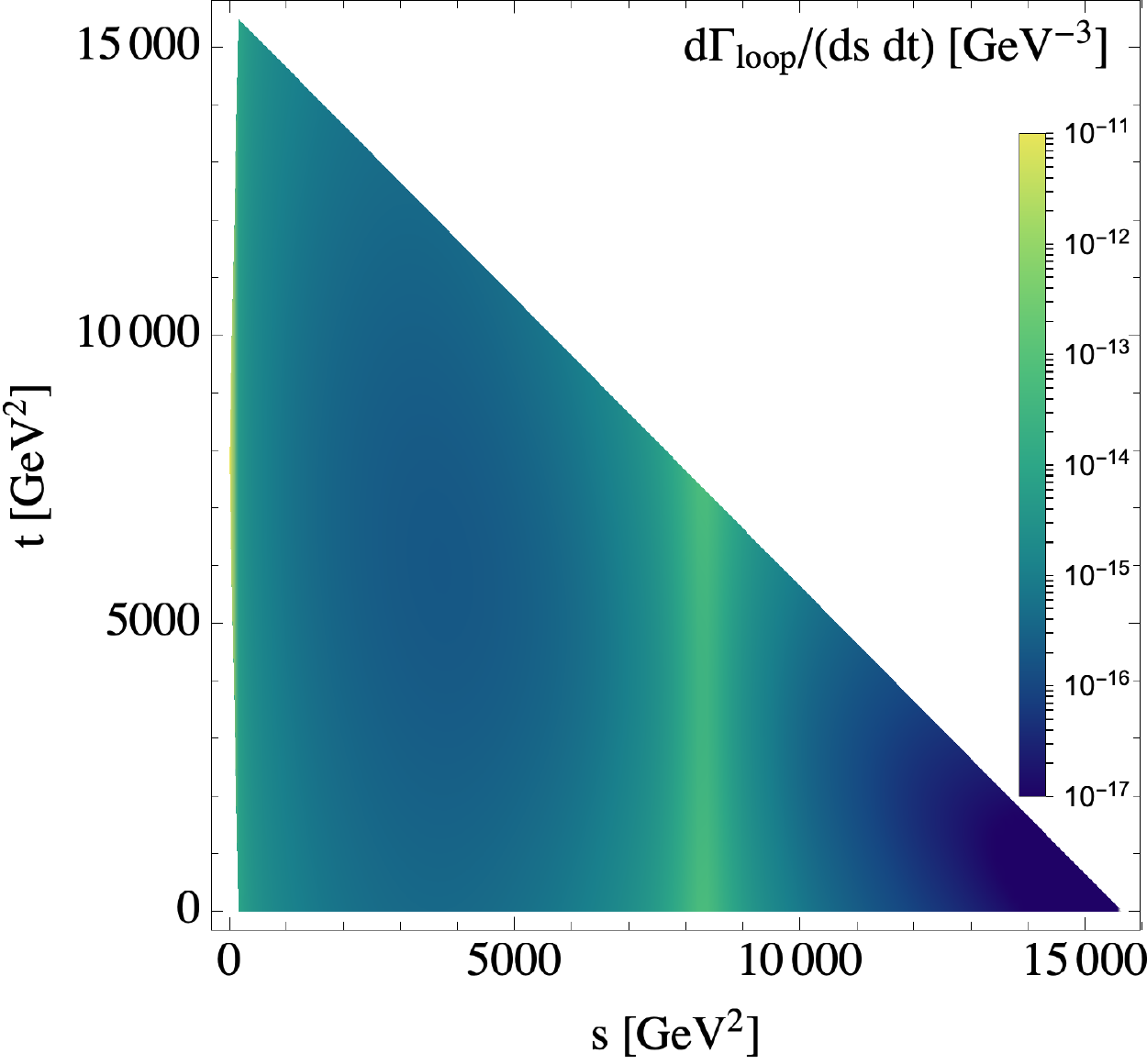}}
		\hspace{0.32cm}
		\subfigure[t][]{\includegraphics[width=0.485\textwidth]{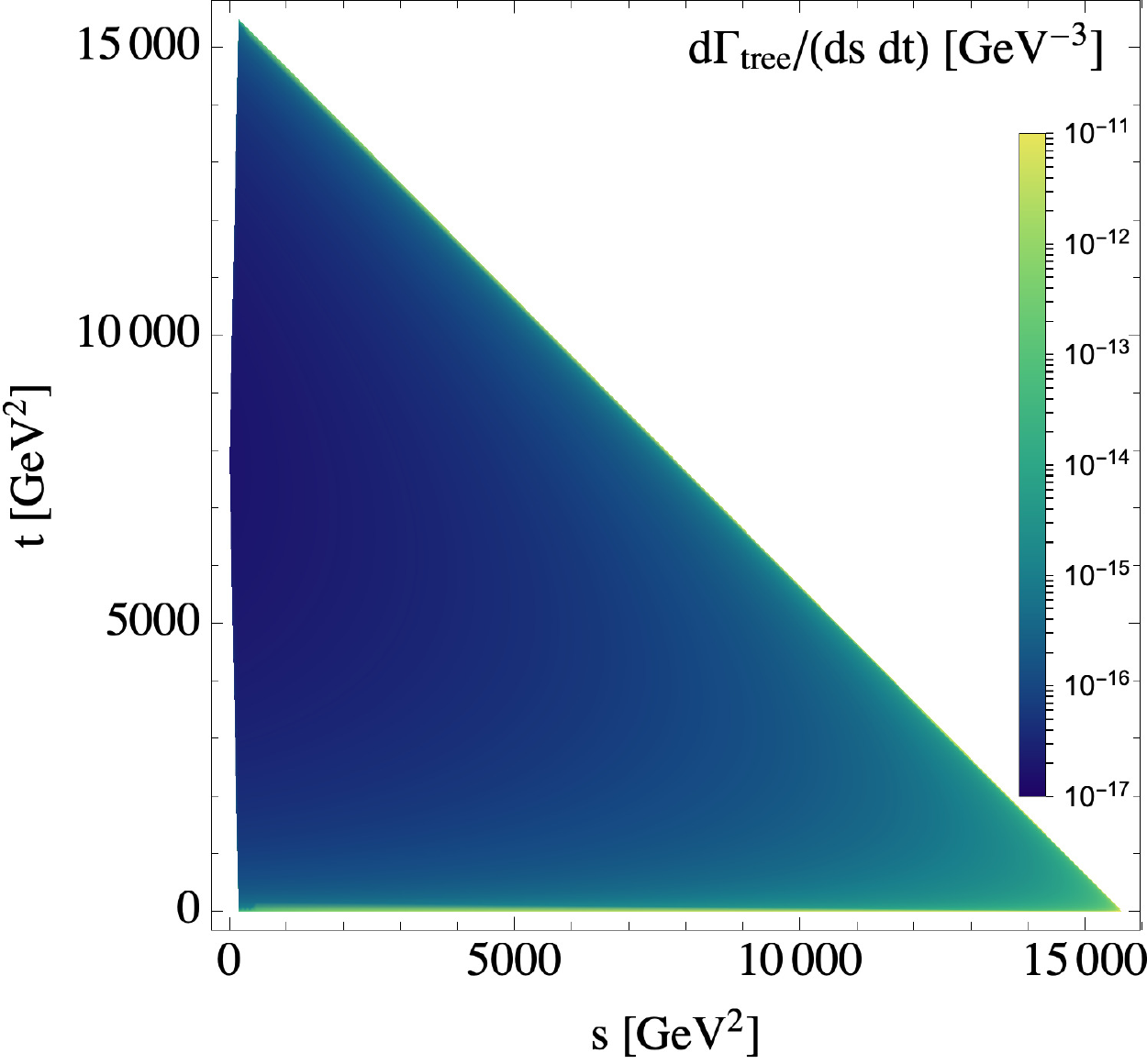}}
         \end{center}
         \caption{Dalitz plot for (a) the one-loop contribution to  the
           decay rate of $h\to \ell^+\ell^-\gamma$
           and (b) the tree contribution to the decay rate of $h\to \mu\mu\gamma$.}
	\label{Fig:Dalitz}
~\\[-3mm]\hrule
\end{figure}

The dependence of the loop- and tree contributions to the differential decay rate on the
Mandelstam variables $s$ and $t$ is displayed in the Dalitz plots in
Fig.~\ref{Fig:Dalitz}. With focus on the kinematic cuts required in the measurements, it is
interesting to observe the behaviour of the distributions in the end-point regions of the
Dalitz plots.  While the one-loop contribution does not increase towards the boundaries, the
tree distribution exhibits strong enhancements in high-$s$, small-$t$, and small $u$
regions, see Eq.~\eqref{Eq:tree-decayrate} above.

With data on $ \frac{d^2\Gamma}{ds\,dt}$ one can implement a very simple
discovery strategy for $H\to Z\gamma$: Just insert $ a_1^{res}$ from
\eq{eq:a1} into \eq{Eq:separating2} and the resulting expression for
$a_1$ into \eq{Eq:squares} (and treat $a_2$ and $b_{1,2}$ in the same
way), then use these results in \eq{loop-decayrate}, and finally add
$ \frac{d^2\Gamma_{\text{tree}}}{ds\,dt}$. When using this formula to
fit the three quantities
$\left[\alpha (m_Z^2)\right]^2 + \left[\beta(m_Z^2)\right]^2$,
$\vert a_1^{nr}\vert ^2+\vert b_1^{nr}\vert ^2$, and
$\vert a_2^{nr}\vert ^2+\vert b_2^{nr}\vert ^2$ to the data, a 5$\sigma$
signal of
$\left[\alpha (m_Z^2)\right]^2 + \left[\beta(m_Z^2)\right]^2\, \neq 0$
will imply the desired discovery. With \eq{Rate:final} below one can
translate this measurement into a number for $\Gamma(H\to
Z\gamma)$. Thus after implementing the lengthy SM expressions for
$a_{1,2}^{nr}$ and $b_{1,2}^{nr}$ one can directly compare
$\Gamma(H\to Z\gamma)$ to the SM prediction in
\eq{Rate:smpred}. 

\begin{figure}[t]
	\begin{center}
		\subfigure[]{\includegraphics[width=0.498\textwidth]{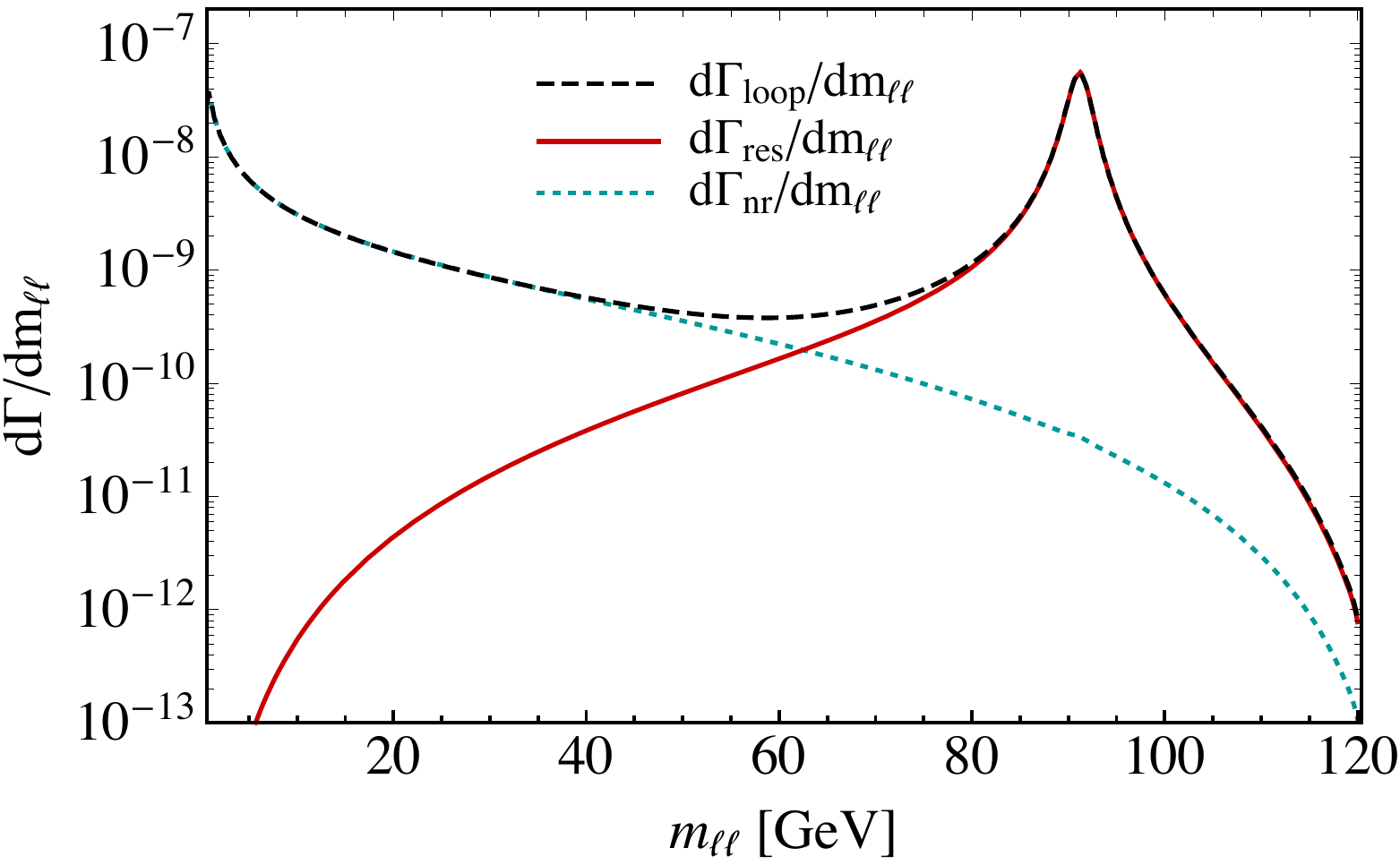}}
		\subfigure[]{\includegraphics[width=0.497\textwidth]{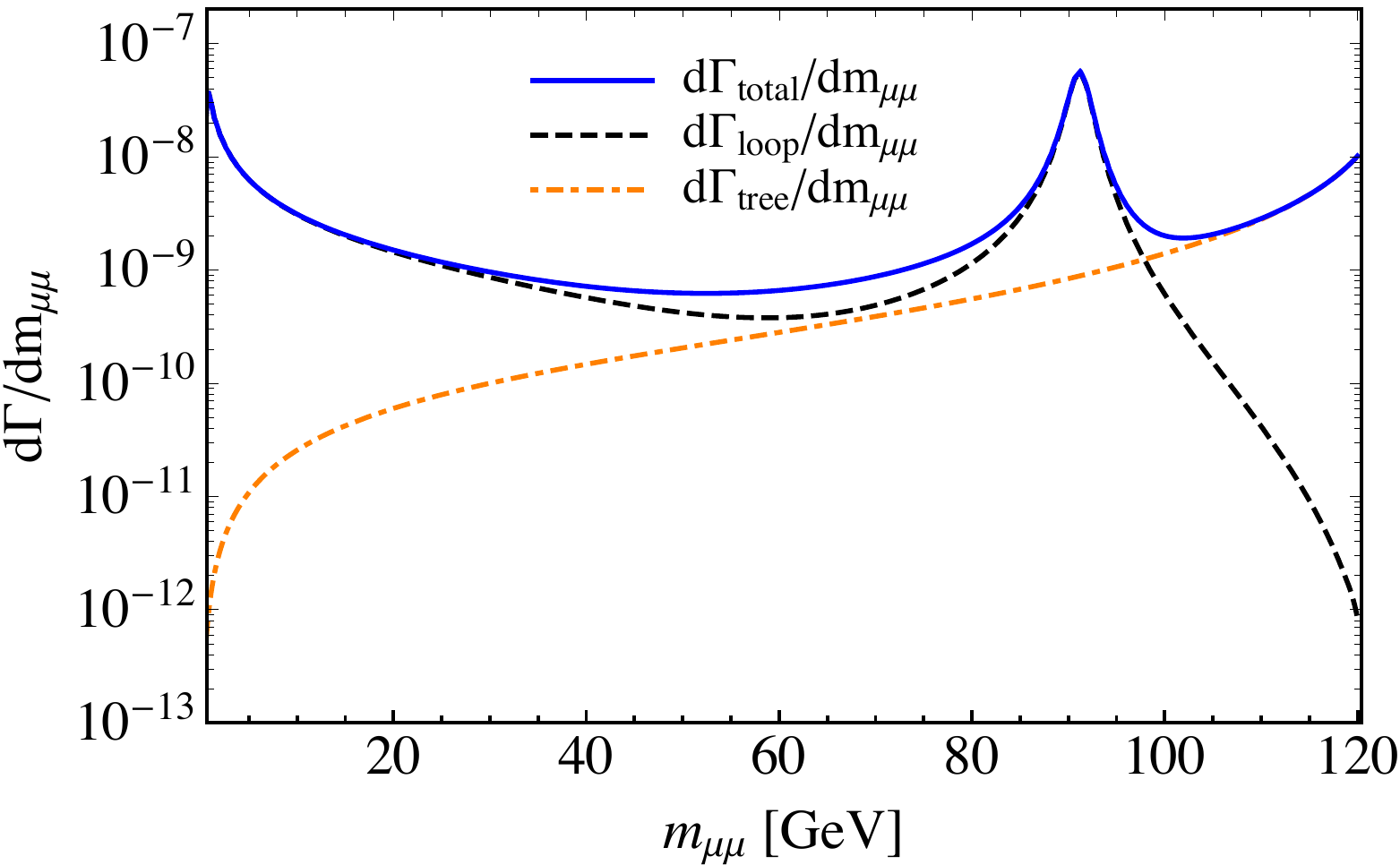}}
         \end{center} 
         \caption{One-loop contributions to differential decay rate with respect to the
           invariant dilepton mass for $\ell = e$ (left) and $\ell=\mu$ (right). The full
           one-loop, resonant and nonresonant contributions are denoted by black dashed,
           solid red and turquoise dot-dashed curves, respectively. For the case $\ell=e$
           the full one-loop contribution represents the full rate, while for $\ell=\mu$,
           the additional, tree-level contribution needs to be accounted for. }
	\label{Fig:Distributions}
~\\[-3mm]\hrule
\end{figure}

Next we discuss the various contributions to $ \frac{d\Gamma}{d m_{\ell\ell}}$, where
$m_{\ell\ell}=\sqrt{s}$ is the dilepton invariant mass. As a first step, we perform the
integration over the  full allowed range of the variable $t$, 
$ t_{\text{min}}\leq t \leq t_{\text{max}}$ with
\begin{eqnarray}
  \displaystyle  t_{\text{min}(\text{max})}(s, m_\ell)
  &=\frac{1}{2}\bigg(m_H^2 - s + 2m_\ell^2\mp (m_H^2 - s)\sqrt{1 - 4 m_\ell^2/s}\bigg)\,.\label{Eq:t-range}
\end{eqnarray}
The resulting resonant and non-resonant one-loop distributions are shown
in the left plot in Fig.~\ref{Fig:Distributions}. Since the masses of
electrons and muons can be safely neglected in the one-loop calculation,
plot (a) represents the loop correction for both cases. Furthermore,
since the tree contribution for $H\to e^+e^-\gamma$ is negligible,
$d\Gamma_{loop}/dm_{\ell\ell}$ also represents the total contribution
for $H\to e^+ e^- \gamma$. The effect of the tree contribution is shown
in the plot~\ref{Fig:Distributions} (b). The only kinematic cut imposed
for these plots is the one for the photon energy in the Higgs rest
frame, $E_{\gamma\text{,\,min}}=\,5\,\text{GeV}$, which only lowers the
maximum value of $m_{\ell\ell}$.

\begin{figure}[t]
	\begin{center}
		\subfigure{\includegraphics[width=0.75\textwidth]{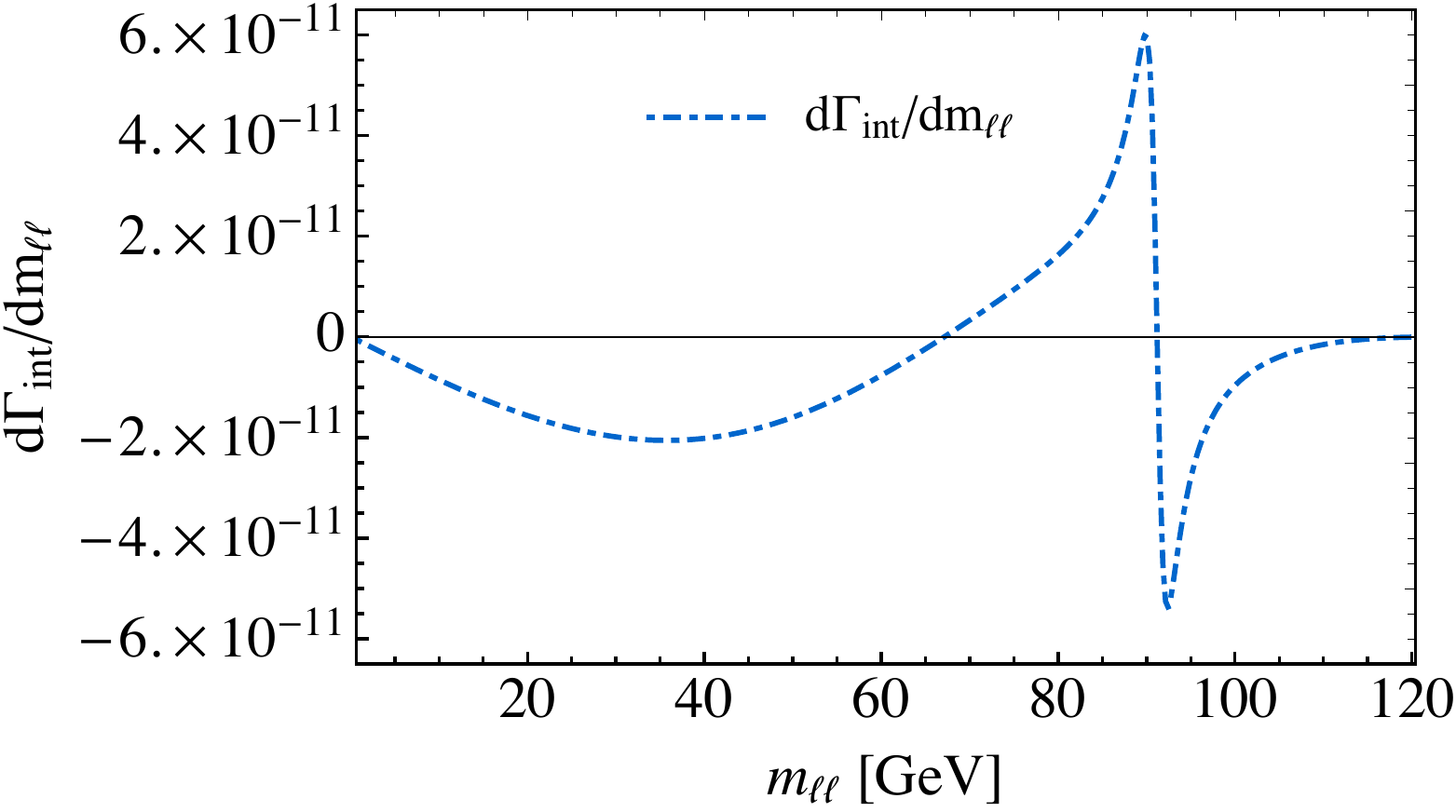}}
         \end{center}
         \caption{Differential distribution $\frac{d\Gamma_{int}}{dm_{\ell\ell}}$ with respect to invariant dilepton
           mass for $\ell = e,\mu$.}
	\label{Fig:Interference}
~\\[-3mm]\hrule
\end{figure}

In Fig.~\ref{Fig:Interference} we display the interference contribution. As expected, this
distribution changes sign at the value of $m_{\ell\ell}$ corresponding to the Z-pole and is
approximately symmetric around the null-axis in this region. However, its magnitude turns
out negligible within the full rate -- this term is completely dropped in the following
discussion.

\section{Kinematic cuts}\label{Sec:Kinematic cuts}
In this section we study the impacts of the kinematic cuts on the minimal values of the
variables $t$ and $u$ on the resonant-, nonresonant- and tree contributions.

We fix the kinematic range for the variable $s$ all the way until the section \ref{Sec:Cuts-Rates} as:
\begin{equation}
\tilde{s}_{min} = (0.1\,m_H)^2\,,\enskip\enskip \tilde{s}_{max} = m_H^2-2\,m_H\, E_{\gamma,min} = (120\, \text{GeV})^2\enskip\,\text{with}\enskip E_{\gamma,min} = 5\,\text{GeV}\,,\label{Eq:s-range1}
\end{equation}
where $E_{\gamma,min}$ the minimal photon energy in the rest frame of the Higgs.

The full physical range for the variable $t$ is given in Eq.~\eqref{Eq:t-range}. We
introduce the kinematic cuts on the minimal values of $t$ and $u$ variables, and denote them
by $\tilde{t}_{min}>0$ and $\tilde{u}_{min}>0$. Note that the cut on the minimal value of
variable $u$ lowers the maximal value of $t$ from the physical limit $t_{max}(s)$ to
$t_{max}(s)-\tilde{u}_{min}$.

Neither the resonant nor the non-resonant loop contribution exhibits a strong
dependence on the small variations of the cuts on the $t,u$-variables near the boundaries of
the Dalitz plot, see Fig.~\ref{Fig:Dalitz} (a) or Eq.~\eqref{Eq:nonresonantCuts} below
On the other hand, the tree contribution is
peaking for the small values of $t$, as can be seen from the Dalitz plot boundary parallel
to $s$-axis, and for the small values of $u$, as can be seen from the diagonal boundary of
the plot in Fig.~\ref{Fig:Dalitz} (b).

\subsection{Resonant contribution}\label{Sec:Resonant contribution}
The resonant distribution is given by:
\begin{equation}
  \displaystyle \frac{d\Gamma_{res}}{ds\,dt}=
  \frac{s (t^2 + u^2)}{512\,m_H^3\pi^3}\frac{1}{(s-m_Z^2)^2+m_Z^2\Gamma_Z^2}
   (\vert \alpha(m_Z^2)\vert^2+\vert \beta(m_Z^2)\vert^2)\,,\label{Eq:Resonant-distrib}
\end{equation}
with the mass of the light lepton neglected in the evaluations of both the kinematics and
the amplitude. With $m_\ell=0$, the physical limits on the variable $t$ are
$t_{min}(s) = 0$, $t_{max}(s) = m_H^2-s$, while $u=m_H^2-s-t$. Numerical values of the loop coefficients at $s=m_Z^2$ are~\cite{Kachanovich:2020xyg}:
\begin{equation}
  \alpha(m_Z^2) =-9.41\cdot 10^{-6}\,\text{GeV}^{-1}\,,\quad\quad
     \beta(m_Z^2) = 1.17\cdot 10^{-5}\,\text{GeV}^{-1}\,.
\end{equation}

Integrating over the variable $t$, while imposing the cuts $\tilde{t}_{min}$ and
$\tilde{u}_{min}$, we have:
\begin{eqnarray}
  \displaystyle \frac{d\Gamma_{res}}{ds} (s, \tilde{t}_{min},\tilde{u}_{min})
  &=&\;
      \frac{s}{512\pi^3 m_H^3} \, \frac{1}{(s-m_Z^2)^2 + m_Z^2 \Gamma_Z^2}
      \, \big(\vert \alpha(m_Z^2)\vert^2
      + \vert \beta(m_Z^2)\vert^2\big)\nonumber\\
  &\cdot& 
          \left[ \frac{t^3 + (s+t-m_H^2)^3}{3}
          \right]_{t=\tilde{t}_{min}}^{t=\tilde{t}_{max}=t_{max}(s)-\tilde{u}_{min}}
          \,.
\label{Eq: dGammads-Res}
\end{eqnarray} 
A further integration over the variable $s$ can also be performed analytically, but results
in a somewhat lengthy expression.
\begin{figure}[t]
	\begin{center}
		\subfigure{\includegraphics[width=0.7\textwidth]{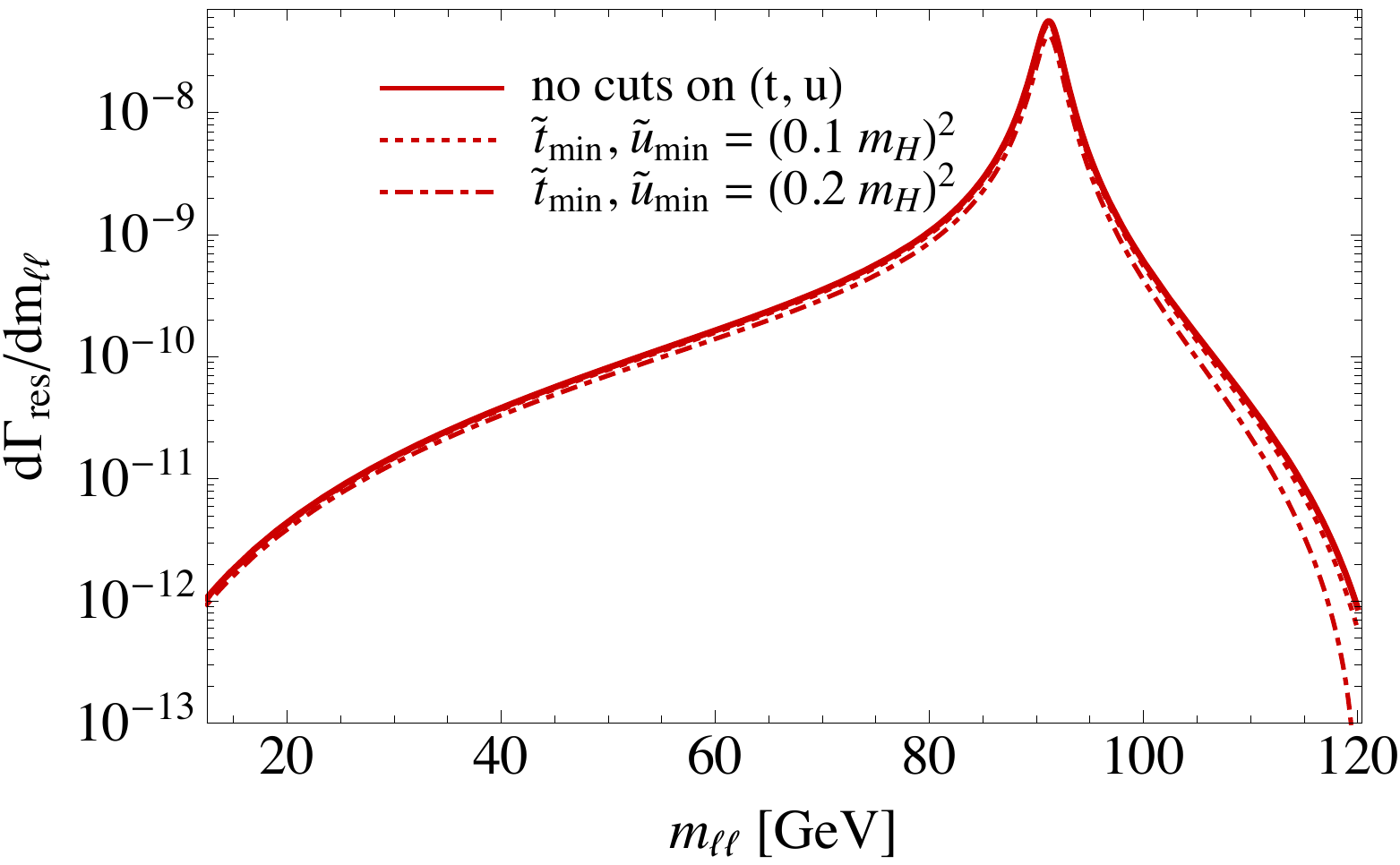}}
		        \end{center}
                        \caption{The resonant decay rate distribution with respect to
                          dilepton invariant mass $m_{\ell\ell}$ for different choices of the cuts
                          $(\tilde{t}_{min}, \tilde{u}_{min})$.}
	\label{Fig:Resonant-tmin}
~\\[-3mm]\hrule
\end{figure}
In Fig.~\ref{Fig:Resonant-tmin} we illustrate the variations of the resonant differential
decay rate $d\Gamma_{res}/dm_{\ell\ell}$ for different values of the cuts
$(\tilde{t}_{min}, \tilde{u}_{min})$.

The effects of the cuts are more noticeable in the fully integrated decay
rate. Integrating over $s$ in the range given in Eq.~\eqref{Eq:s-range1} we have, e.g.
\begin{eqnarray}
\displaystyle\frac{\Gamma_{res}[\tilde{t}_{min}=(\kappa\,m_H)^2, \tilde{u}_{min}=(\kappa\,m_H)^2]}{\Gamma_{res}[\tilde{t}_{min}=0, \tilde{u}_{min}=0]} = (1, 0.94, 0.77)\,,\enskip\enskip\enskip \text{for}\enskip\enskip \kappa = (0\,,0.1\,,0.2)\,,
\end{eqnarray}
with
\begin{equation}
\Gamma_{res}[\tilde{t}_{min}=0,\tilde{u}_{min}=0]=0.215 \, \text{keV}\,.
\end{equation}

\subsection{Nonresonant contribution}\label{Sec:Nonresonant contribution}
\begin{figure}[t]
	\begin{center}
		\subfigure{\includegraphics[width=0.7\textwidth]{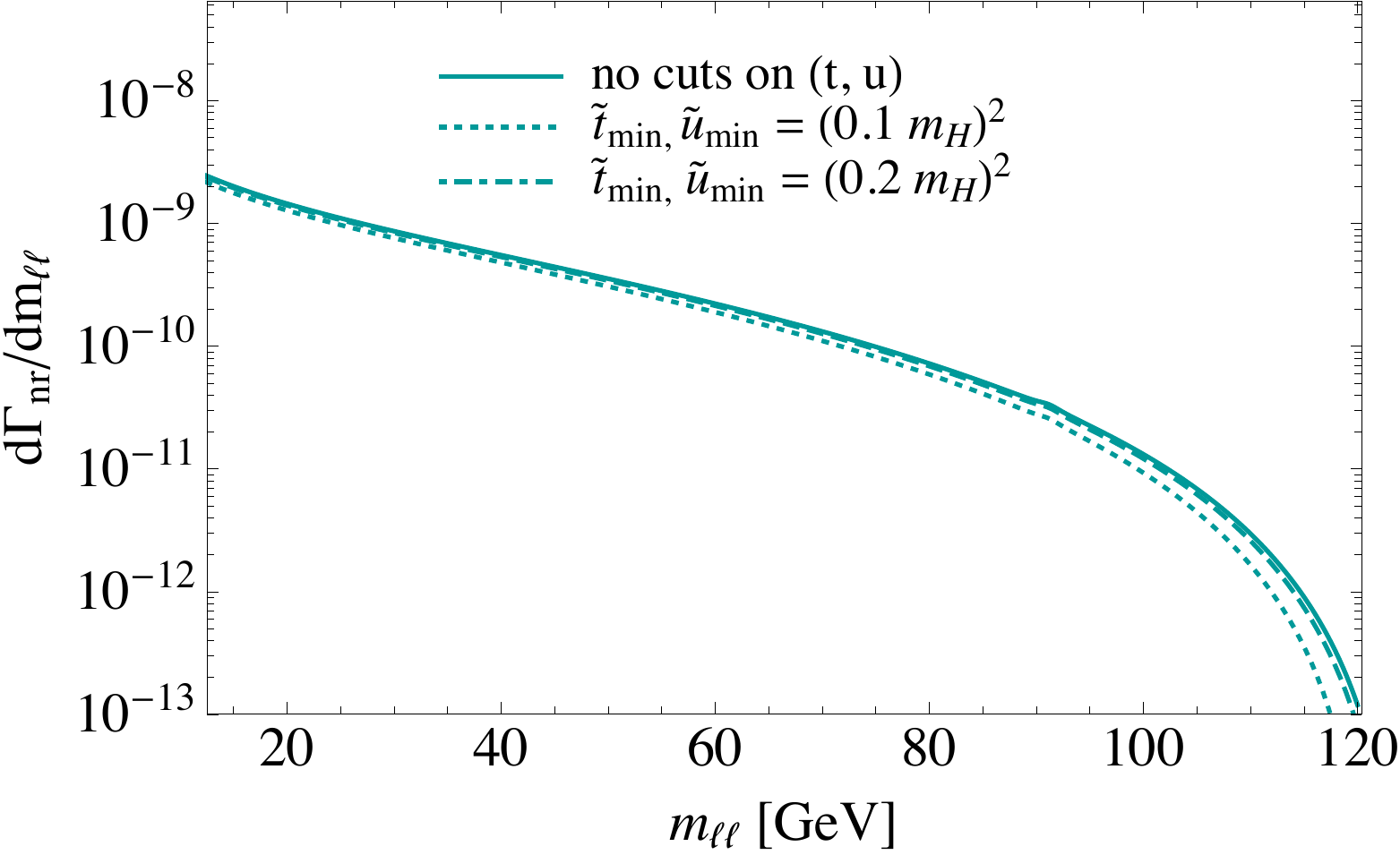}}
		        \end{center}
         \caption{The nonresonant decay distributions $d\Gamma_{nr}/dm_{\ell\ell}$, for few choices of the cut $\tilde{t}_{min}$.}
	\label{Fig:Nonresonant-tmin}
~\\[-3mm]\hrule
\end{figure}

The analytic form of the non-resonant contribution turns out rather lengthy -- its explicit
form can be read off from the expressions given in Appendix A of
Ref.~\cite{Kachanovich:2020xyg}. As in the previous case, we integrate the corresponding
decay distribution over the variable $t$ numerically from $\tilde{t}_{min}>0$ to the value
$t_{max}(s)-\tilde{u}_{min} = m_H^2-s-\tilde{u}_{min}$.  We illustrate the effect of several
choices of the cuts $\tilde{t}_{min}\,,\tilde{u}_{min}$ on the nonresonant differential
distribution over $m_{\ell\ell}$ in Fig.~\ref{Fig:Nonresonant-tmin}.
Again, integrating over the variable $s$ in the limits given in  Eq.~\eqref{Eq:s-range1}, we obtain:
\begin{eqnarray}
\displaystyle\frac{\Gamma_{nr}[\tilde{t}_{min}=(\kappa\,m_H)^2, \tilde{u}_{min}=(\kappa\,m_H)^2]}{\Gamma_{nr}[\tilde{t}_{min}=0, \tilde{u}_{min}=0]} = (1, 0.97, 0.87)\,,\enskip\enskip\enskip \text{for}\enskip\enskip \kappa = (0\,,0.1\,,0.2)\,,
\label{Eq:nonresonantCuts}
\end{eqnarray}
where
\begin{equation}
\Gamma_{nr}[\tilde{t}_{min}=0, \tilde{u}_{min}=0]=0.043 \, \text{keV}\,.
\end{equation}
Therefore, we find weak dependence on the $t,u$-cuts as long as the values of the latter are not such that they remove a significant amount of the phase space.

It is convenient to display the shapes of the distributions shown in Fig.~\ref{Fig:Nonresonant-tmin} in an approximate numerical form. Since the dependence on the cuts is small, we represent the shape that does not involve any cuts on variables $t,u$ as the following power series:
\begin{equation}
\frac{d\Gamma_{nr}}{dm_{\ell\ell}} = 10^{-10}\sum_{n=-4}^3 c_n \bigg(\frac{m_{\ell\ell}}{m_H}\bigg)^n\,+\ldots\label{Eq:NRApprox}
\end{equation}
with
\begin{equation}
(c_{-4}, \ldots,  c_{3}) = (3.27\cdot 10^{-4}\,, -1.26\cdot 10^{-2}\,, 2.0\cdot 10^{-1}\,, 8.49\cdot 10^{-1}\,, 7.96\,, -30.1\,, 32.1\,, -11.0)\,.
\end{equation}
The integral of the above approximate function over the variable $m_{\ell\ell}$ differs from the exact result at the level of around $0.5\%\,(2\%)$ for $m_{\ell\ell\,,min}=0.1\,m_H\enskip(0.5 m_H)$, with $m_{\ell\ell, max}=120$ GeV for both cases. This is an acceptable approximation given that the non-resonant part is itself a small contribution to the full decay rate in the interesting region around $Z$-boson peak. 

\subsection{Tree contribution}\label{Sec:Tree contribution}
The definite integral over the variable $t$ in \eq{Eq:tree-decayrate} can be performed
analytically. As before, for the lower limit we have $\tilde{t}_{min}$, which is larger or
equal to the the physical lower limit $t_{min}(s,m_\ell)$, while the upper limit is
$t_{max}(s,m_\ell)-\tilde{u}_{min}$. Introducing the shorthand notation%
\beq%
\mathcal{I}(t)= \int dt\, \frac{d^2\Gamma_{tree}}{ds\,dt},\quad\quad\quad
\mathcal{I}(a,b)\equiv\mathcal{I}(b)-\mathcal{I}(a)\,, %
\eeq%
the resulting distribution with respect to $s$ is:
\begin{equation}
\begin{split}
\frac{d\Gamma_{tree}}{ds}(s;\,\tilde{t}_{min}\, \tilde{u}_{min}) &= \int_{\tilde{t}_{min}}^{t_{max}-\tilde{u}_{min}}dt\,\frac{d^2\Gamma_{tree}}{ds\, dt}\theta\big(t-t_{min}(s)\big)\\
& =\mathcal{I}\big(t_{min}(s),t_{max}(s)-\tilde{u}_{min}\big)-\theta\big(\tilde{t}_{min}-t_{min}(s)\big)\mathcal{I}\big(t_{min}(s),\tilde{t}_{min}\big)\,,\label{Eq:int-tree}
\end{split} 
\end{equation}
where we have temporarily suppressed an additional dependence of $t_{min(max)}$ on the
lepton mass, for clarity of the notation. Note that the insertions of the Heaviside step
function in the above equation confine the integration to the physically allowed
region. {The expression for $\mathcal{I}(t)$ is:}
\beq
\begin{split}
  \mathcal{I}(t) = &\frac{\alpha^2 m_\ell^2}{16\,\pi\, m_H^3 m_W^2 \sin^2\theta_W}\bigg[\frac{2m_\ell^2(m_H^2-4m_\ell^2)}{t-m_\ell^2}+\frac{2m_\ell^2(m_H^2-4 m_\ell^2)}{s + t - m_H^2 - m_\ell^2}\\
  & \qquad\qquad - \frac{m_H^4-4 m_H^2 m_\ell^2 +
    (s-4m_\ell^2)^2}{s-m_H^2}\ln\bigg(\frac{s+t-m_H^2-m_\ell^2}{t-m_\ell^2}\bigg)\bigg]\,.
\end{split}\label{Eq:int-tree2}
\eeq
The final formula for
$\frac{d^2\Gamma_{tree}}{ds}(s;\,\tilde{t}_{min}, \tilde{u}_{min})$ is obtained by inserting
the result of \eq{Eq:int-tree2} into Eq.~\eqref{Eq:int-tree}.
\begin{figure}[t]
	\begin{center}
		\subfigure{\includegraphics[width=0.7\textwidth]{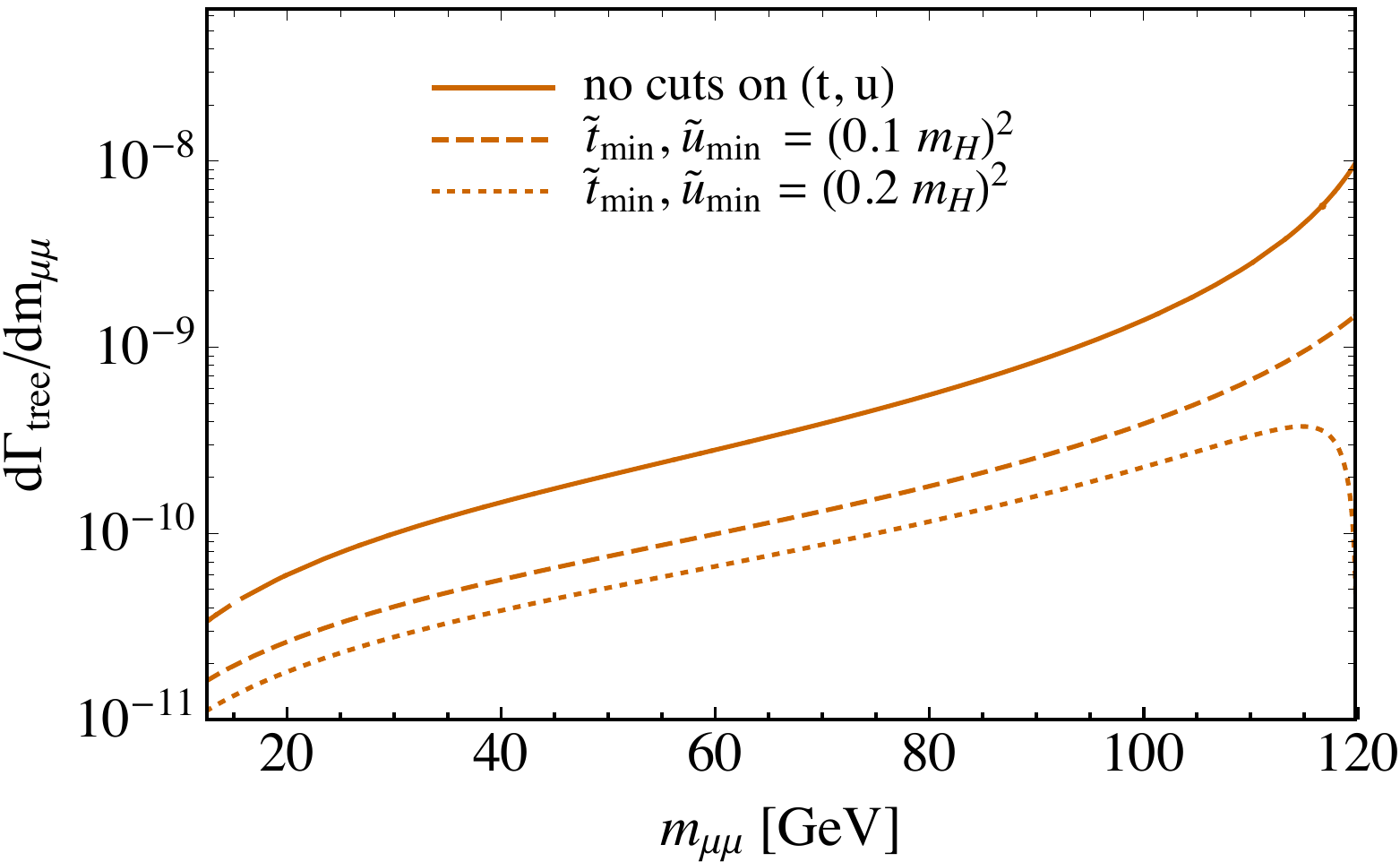}}
         \end{center}
         \caption{Differential distribution $d\Gamma_{tree}/dm_{\mu\mu}$ with respect to invariant dimuon
           mass.}
	\label{Fig:Tree-tmin}
~\\[-3mm]\hrule
\end{figure}
We illustrate the dependence of the tree contribution on the cuts for several values of
$\tilde{t}_{min}$ and $\tilde{u}_{min}$ in Fig.~\ref{Fig:Tree-tmin}.

Finally, integrating over the variable $s$ in the limits given in Eq.~\eqref{Eq:s-range1}, we have:
\begin{eqnarray}
\displaystyle\frac{\Gamma_{tree}[\tilde{t}_{min} = (\kappa\,m_H)^2, {\tilde{u}_{min} = (\kappa\,m_H)^2}]}{\Gamma_{tree}[t_{min}(s,m_\mu)]} = (1,\,0.25,\,0.12)\,,\enskip\enskip\enskip \text{for}\enskip\enskip \kappa = (0\,,0.1\,,0.2)\,,
\label{Eq:Gamma-tree}
\end{eqnarray}
where
\begin{equation}
\Gamma_{tree}[\tilde{t}_{min}=t_{min}(s,m_\mu)]=0.104 \, \text{keV}\,.
\end{equation}

\subsection{Kinematic cuts and total rates}\label{Sec:Cuts-Rates}
We now explore how each of the three contributions to integrated decay rate depends on the
cuts on variables that also include $s$. We propose different cuts to optimize the
sensitivity to the three milestones mentioned in the abstract.  
The results for several combinations of such cuts
are shown in Table~\ref{Tab:cutss}.

Cuts 1 and 2 correspond to the choices of the three previous subsections\footnote{The upper
  limit on $s=(120\,\text{GeV})^2$, set for these two cuts, is the result of imposing a
  minimal photon energy, see Eq.~\eqref{Eq:s-range1}.}. For the cut 1 we find that the
nonresonant contribution is around $20\%$ of the resonant one, while the tree contribution
is somewhat larger than about $10\%$. As noted before, the tree contribution receives a
strong suppression with the increasing vales of $\tilde{t}_{min}$ and $\tilde{u}_{min}$.
Cuts 3 and 4 isolate the resonant contribution stemming from $H\to Z\gamma$, while cuts
5 and 6 probe the nonresonant contribution. The purpose of cut 7 is the isolation of the tree
contribution. Cut 8 simply illustrates an additional suppression of the tree contribution
that results from tightening of cuts on $t$ and $u$.
\begin{table}[t]
  \caption{\label{Tab:cutss}Integrated decay rates for different
    contributions to $H\to \mu^+\mu^-\gamma$ for several choices of the
    kinematic cuts on the variables $s, t$ and $u$. Note the symmetric
    choice  $\tilde{u}_{min}=\tilde{t}_{min}$.}
\begin{center}
\begin{tabular}{ |p{1.cm}||p{1.8cm}|p{1.8cm}|p{1.8cm}|p{1.8cm}|p{1.8cm}|p{1.8cm}|p{1.8cm}||p{1.8cm}|}
 \hline
 Cut &$s_{min}$&$s_{max}$&$\tilde{t}_{min},\tilde{u}_{min}$&$\Gamma_{res}$(keV)&$\Gamma_{nr}$(keV)&$\Gamma_{tree}$(keV)&$\Gamma_{tot}$(keV)&Purpose\\
 \hline
 \hline
 1&$(0.1\, m_H)^2$&$(120\,\text{GeV})^2$&$(0.1\, m_H)^2$&$0.202$&$0.042$&$0.026$&$0.270$&general\\
 \hline
  2 &$(0.1\, m_H)^2$&$(120\,\text{GeV})^2$&$(0.2\, m_H)^2$&$0.165$&$0.037$&$0.013$&$0.215$&general\\
  \hline
  \hline
  3 &$(70\,\text{GeV})^2$&$(100\,\text{GeV})^2$&$(0.1\, m_H)^2$&$0.195$&$0.002$&$0.007$&$0.204$&$h\to Z\gamma$\\
  \hline
  4 &$(70\,\text{GeV})^2$&$(100\,\text{GeV})^2$&$(0.2\, m_H)^2$&$0.160$&$0.001$&$0.004$&$0.165$&$h\to Z\gamma$\\
  \hline
  \hline
  5 &$(10\,\text{GeV})^2$&$(40\,\text{GeV})^2$&$(0.1\, m_H)^2$&$3.53\cdot 10^{-4}$&$3.78\cdot 10^{-2}$&$1.02\cdot 10^{-3}$&$3.92\cdot 10^{-2}$& nonresonant\\
  \hline
  6 &$(20\,\text{GeV})^2$&$(40\,\text{GeV})^2$&$(0.1\, m_H)^2$&$3.33\cdot 10^{-4}$&$1.75\cdot 10^{-2}$&$8.12\cdot 10^{-4}$&$1.87\cdot 10^{-2}$& nonresonant\\
  \hline
  \hline
  7 &$(100\,\text{GeV})^2$&$(120\,\text{GeV})^2$&$(0.1\, m_H)^2$&$1.93\cdot 10^{-3}$&$7.51\cdot 10^{-5}$&$1.5\cdot 10^{-2}$&$1.70\cdot 10^{-2}$& tree\\
  \hline
  8 &$(100\,\text{GeV})^2$&$(120\,\text{GeV})^2$&$(0.2\, m_H)^2$&$1.40\cdot 10^{-3}$&$5.28\cdot 10^{-5}$&$6.06\cdot 10^{-3}$&$7.51\cdot 10^{-3}$ &tree\\
  \hline
\end{tabular}
\end{center}
\end{table}

\section{Resonant contribution and the Narrow-Width Approximation\label{Sec:nwa}}
The resonant contribution is related to the decay rate of $H\to Z\gamma$ involving an
on-shell $Z$ boson that subsequently decays to a pair of light leptons.

We recall the amplitude for the process $H\to Z\gamma$:
\begin{equation}
\mathcal{A}=\mathcal{ \widetilde{A}}\,\bigg[ \big(p_Z\cdot \epsilon(q)^\ast\big)\big(q\cdot \epsilon(p_Z)^\ast\big)-\big(p_Z\cdot q\big)\big(\epsilon(q)^\ast\cdot \epsilon(p_Z)^\ast\big)\bigg]\,,\label{Eq:htoZgamma}
\end{equation}
where $p_Z, q,  \epsilon(p_Z), \epsilon(q)$ denote momenta and polarizations of $Z$-boson and photon, respectively, while the loop function $\tilde{\mathcal{A}}$ is given in Eq.~\eqref{Eq:Atilde}.
The decay rate is:
\begin{equation}
\Gamma({H\to Z\gamma}) = \frac{(m_H^2-m_Z^2)^3}{32\pi m_H^3}\vert \mathcal{\widetilde{A}}\vert^2\,.\label{Rate:HtoZgamma}
\end{equation} 
in agreement with the result in Ref. \cite{Djouadi:2005gi}.
Evaluating $\tilde{\mathcal{A}}$ in \eq{Eq:Atilde} for the input
values of \eq{inputs} gives the SM prediction
\begin{equation}
  \Gamma({H\to Z\gamma}) =6.51\,\text{keV}, 
  \label{Rate:smpred}
\end{equation}
again in agreement with the numerical result found from the analytic
expression in Ref.~\cite{Djouadi:2005gi}.  This value is $3\%$ larger
than the central value quoted by the LHC Higgs Cross Section Working
Group, $\Gamma({H\to Z\gamma}) =6.31\,\text{keV}$, in Table~177 on page
679 of Ref.~\cite{deFlorian:2016spz}, see also Eq. (III.1.18) on page
403. Ref.~\cite{deFlorian:2016spz} finds an uncertainty of the theory
prediction of order $5\%$, which could be reduced by a two-loop
calculation.

Furthermore, the branching ratio of the process $Z \to \ell\ell$ at
tree-level is
\begin{equation}
  BR(Z \to \ell\ell) = \; \frac{m_Z}{\Gamma_Z}  \widetilde{C}\,, \qquad\qquad
  \widetilde{C} =\; \frac{e^2 (8 \sin^4 \theta_W - 4 \sin^2 \theta_W +1)}{96\pi \cos^2\theta_W
    \sin^2\theta_W} \stackrel{\text{Eq.}(\ref{inputs})}{=} 9.2\cdot 10^{-4}
  \,.\label{Rate: Ztoll}
\end{equation}
Integration of the resonant distribution $d^2 \Gamma_{res}/(ds\,dt)$ over the variable $t$ in the full range given in Eq.~\eqref{Eq:t-range} results in
\begin{eqnarray}
\frac{d\Gamma_{res}}{ds}=\frac{s}{512\pi^3 m_H^3}\frac{1}{(s-m_Z^2)^2 + m_Z^2 \Gamma_Z^2}\cdot\frac{2}{3} (m_H^2-s)^3 \cdot \bigg(\vert \alpha(m_Z^2)\vert^2 + \vert \beta(m_Z^2)\vert^2\bigg)\,.\label{Eq: dGammads-Res2}
\end{eqnarray}
We now apply the narrow-width approximation (NWA) for the Breit-Wigner distribution: 
\begin{equation}
\text{NWA}:\enskip\enskip \frac{\Gamma_Z}{m_Z}\to 0\,, \quad\quad \frac{1}{(s-m_Z^2)^2 + m_Z^2 \Gamma_Z^2} \to \frac{\pi}{m_Z \Gamma_Z} \delta(s-m_Z^2)\,,\label{NWA}
\end{equation}
where the limit is taken under the integral over $s$.  Substituting this limit into
Eq.~\eqref{Eq: dGammads-Res2}, integrating this distribution over
$s$, and using the relations \eqref{Rate:HtoZgamma} and \eqref{Rate: Ztoll} we find:
\begin{equation}
\Gamma_{\text{NWA}}= \Gamma(H\to Z\gamma)\cdot BR(Z\to \ell\ell)\,,
\end{equation} 
provided that
\begin{equation}
   \left[\alpha(m_Z^2)\right]^2 + \left[\beta(m_Z^2)\right]^2 = 24 \pi
   \widetilde{\mathcal{A}}^2 \widetilde{C}\,.\label{The-relation}
\end{equation}
The latter relation can be explicitly confirmed using the functions $\alpha(s)$ and
$\beta(s)$, given in Eqs. A.1 and A.2 in Ref.\cite{Kachanovich:2020xyg}.
Thus if $ \left[\alpha (m_Z^2)\right]^2 +  \left[\beta(m_Z^2)\right]^2 $ extracted from data, the desired decay width is
calculated as
\begin{equation}
  \Gamma({H\to Z\gamma}) \; =\;\;
  \frac{(m_H^2-m_Z^2)^3}{32\pi m_H^3} \;
    \frac{\left[\alpha (m_Z^2)\right]^2 +  \left[\beta(m_Z^2)\right]^2 }{24\pi  \widetilde{C}}
    \; \stackrel{\text{Eq.}(\ref{inputs})}{=} \; (30.687\,\text{GeV})^3 \, \times \,
         \left[ \left[\alpha (m_Z^2)\right]^2 +  \left[\beta(m_Z^2)\right]^2\right] 
    \label{Rate:final}
\end{equation}
with $\widetilde{C}$ defined in \eq{Rate: Ztoll}. 

Using \eq{The-relation} we can rewrite Eq.~\eqref{Eq: dGammads-Res} as
\begin{eqnarray}
\displaystyle \frac{d\Gamma_{res}}{ds} (s, \tilde{t}_{min},\tilde{u}_{min})
  &=&\Gamma(H\to Z\gamma)\cdot BR(Z\to
      \ell\ell) \cdot
      \frac{3\,s\,\Gamma_Z}{2\pi m_Z (m_H^2-m_Z^2)^3}  \nonumber\\
  && \quad \cdot \frac{1}{(s-m_Z^2)^2 + m_Z^2
     \Gamma_Z^2} \, \left[   \frac{t^3 + (s+t-m_H^2)^3}{3}  \right]_{t=\tilde{t}_{min}}^{t=t_{max}(s)-\tilde{u}_{min}}\,.
\label{Eq: dGammads-Resv2}
\end{eqnarray} 
The resulting decay rate is expressed as the function of the kinematic cuts
$\tilde{t}_{min}$, $\tilde{u}_{min}$ and can be readily compared to the leading order result for
\begin{equation}
  \Gamma_{NWA} = \Gamma (H\to Z\gamma)\cdot BR(Z\to \ell\ell) = 0.219\,\text{keV}\,
   =0.0336 \, \times \,  \Gamma (H\to Z\gamma)
\end{equation}
  obtained
using the parameter inputs from Eq.~\eqref{inputs}.

\section{Conclusions\label{Sec:con}}
The decay rate
  $ \frac{d\Gamma (H\to \ell^+\ell^- \gamma)}{d m_{\ell\ell}}$ with
  $\ell=e$ or $\mu$ offers insights into different aspects of 
  Higgs physics. With increasing integrated luminosity it will be
  possible to (i) discover the decay $H\to Z \gamma$ and measure its
  branching ratio, (ii) discover the decay
  $H\to \mu^+\mu^- \left. \gamma\right|_{\rm tree}$ driven by the muon
  Yukawa coupling, and (iii) ultimately quantify potential new physics
  contributions to both the loop-induced $H\to Z \gamma$ decay and the
  off-peak contributions to $H\to \ell^+\ell^- \gamma$. The latter
  comprise the non-resonant loop contributions, best tested in the
  region between the photon and $Z$ poles, and (for $\ell=\mu$)
  $H\to \mu^+\mu^- \left. \gamma\right|_{\rm tree}$ which dominates
  $ \frac{d\Gamma (H\to \ell^+\ell^- \gamma)}{d m_{\ell\ell}}$ near the
  endpoint region with $ m_{\ell\ell}> M_Z$.

  In this paper we have proposed a gauge-independent, physical
  definition of the decay rate $\Gamma(H\to \ell^+\ell^- \gamma)$ and
  shown how it can be extracted from the measured decay spectrum
  $ \frac{d\Gamma (H\to \ell^+\ell^- \gamma)}{d m_{\ell\ell}}$. To this
  end it is necessary to subtract the non-resonant contribution to
  $ \frac{d\Gamma (H\to \ell^+\ell^- \gamma)}{d m_{\ell\ell}}$ and we
  have derived easy-to-use approximations for the cumbersome SM
  expression, see \eq{Eq:NRApprox} above.  We have further studied the
  dependence of
  $ \frac{d\Gamma (H\to \ell^+\ell^- \gamma)}{d m_{\ell\ell}}$ on
  kinematical cuts, which we only found to be a critical issue for
  $H\to \mu^+\mu^- \left. \gamma\right|_{\rm tree}$. In order to perform
  the three milestone measurements mentioned above we have proposed cuts
  to optimize the sensitivities to $H\to Z \gamma$,
  $H\to \mu^+\mu^- \left. \gamma\right|_{\rm tree}$, and the
  non-resonant loop contribution, respectively, see
  Table~\ref{Tab:cutss}. 

\section*{Acknowledgments}
A.K.\ and U.N.\ acknowledge support by DFG through
CRC TRR 257, \emph{Particle Physics Phenomenology after the Higgs
  Discovery} (grant no.~396021762). I.N. would like to acknowledge support from the Alexander von Humboldt Foundation within the Research Group Linkage
Programme funded by the German Federal Ministry of Education and
Research.

\appendix
\section{Inputs}
We use the following values for the parameter inputs:  
\begin{equation}
\begin{split}
  &\qquad  m_W = 80.379\,
  \text{GeV}\,,\qquad m_Z = 91.1876\,\text{GeV}\,,\qquad \sin^2\theta_W = 1-\frac{m_W^2}{m_Z^2}=0.223013\,,\\
&  \qquad\qquad 
  m_t = 173.1\,\text{GeV} \,,\qquad m_H = 125.1\,\text{GeV}\,,\qquad m_\mu=0.105658\,\text{GeV}\,,\qquad \Gamma_Z = 2.4952\,\text{GeV}\,,\\\
  & G_F=1.1663787\times 10^{-5}\,\text{GeV}^{-2}\,,\qquad \alpha^{-1} =
  \frac{\pi}{\sqrt{2}G_F m_W^2 \sin^2\theta_W} = 132.184\,.
\label{inputs}
\end{split}
\end{equation}

\section{The loop function $\tilde{\mathcal{A}}$}\label{Atilde}
The loop function $\tilde{\mathcal{A}}$, introduced in Eq.~\eqref{Eq:htoZgamma}, is given as:
\begin{eqnarray}
\nonumber \tilde{\mathcal{A}}  &=& \frac{e^3}{3 \cdot 16 \pi^2 \cos \theta_{W} \sin^{2} \theta_{W} m_{W}^{2}(m_{H}^{2} - m_{Z}^{2})^{2}} \times \Big\{4 \left(5 - 8 \cos^{2}\theta_{W}\right) m_{t}^{2} m_{Z}^{2}  m_{W} \\
\nonumber && \times (B_{0}(m_{H}^{2}, m_{t}^{2}, m_{t}^{2}) - B_{0}(m_{Z}^{2}, m_{t}^{2}, m_{t}^{2})) \\
\nonumber &-& 3 m_{W} m_{Z}^{2} \left(2 m_{W}^{2} + m_{H}^{2} - 12 \cos^{2}\theta_{W} m_{W}^2 - 2 \cos^{2}\theta_{W} m_{H}^2 \right) \\
\nonumber  && \times  \left(B_{0}(m_{H}^{2}, m_{W}^{2}, m_{W}^{2}) - B_{0}(m_{Z}^{2}, m_{W}^{2}, m_{W}^{2}) \right) \\
\nonumber &+& m_{W}(m_{Z}^{2} - m_{H}^{2}) \Big(  2 (5 - 8 \cos^{2}\theta_{W})m_t^2  \\
\nonumber  &&\times (m_H^2 -4 m_t^2  - m_Z^2) C_{0} (0, m_{H}^{2}, m_{Z}^{2}, m_{t}^{2}, m_{t}^{2}, m_{t}^{2}) \\
\nonumber &-& 6 m_{W}^{2} \left(\big(1 - 6\cos^{2}\theta_{W}\big) m_{H}^{2} + 2(6\cos^{4}\theta_{W} + 3 \cos^{2}\theta_{W}- 1)m_{Z}^{2} \right) \\
\nonumber && \times C_{0}(0, m_{H}^{2}, m_{Z}^{2}, m_{W}^{2}, m_{W}^{2}, m_{W}^{2})  \\
&+& (3 - 6  \cos^{2}\theta_{W} ) m_{H}^{2} + 4 (8 \cos^{2}\theta_{W}- 5 ) m_{t}^{2} + 6 (1- 6 \cos^{2}\theta_{W}) m_{W}^{2} \Big)  \Big\}\,,
\label{Eq:Atilde}
\end{eqnarray}
expressed in terms of Veltman-Passarino loop functions~\cite{Passarino:1978jh}, following the conventions of \texttt{Feyncalc}~\cite{Shtabovenko:2020gxv,Shtabovenko:2016sxi,Mertig:1990an} package.

\end{document}